\documentclass{evn2004}
\usepackage{txfonts}
\begin{document}
   \title{What is the primary beam response of an interferometer with unequal 
elements?}

   \author{Richard G. Strom\inst{1,}\inst{2}
          }

   \institute{ASTRON, Radio Observatory, Postbus 2, 7990 AA Dwingeloo,
The Netherlands
         \and
             Astronomical Institute, University of Amsterdam, Kruislaan 403,
1098 SJ Amsterdam, The Netherlands
             }

   \abstract{The EVN stations encompass elements with a range of diameters,
even including an interferometer (the Westerbork Telescope, with up to 14
elements used together as a tied array). In combination, the various
station pairs will each produce their own primary beam envelopes, with
which the interferometer pattern is modulated. People sometimes forget
that in the case of unequal elements, this combined primary beam envelope
is different from the beam of each element separately. The reason for
this is reviewed, the results for a number of station pairs are
summarized, and some of the practical consequences are discussed. The
increased interest in wide-field applications, as illustrated by several
recent results, underlines the need for a proper determination of the
interferometer beam envelope.
   }

   \maketitle
%

\section{Introduction}

This short note is motivated by comments I have occasionally heard (or
read in proposals or similar unpublished documents), in the hope that 
it will correct a misconception some people seem to have. The
argument goes something like this: We want an in-beam phase reference.
Hmmm, our largest element will be Effelsberg, it has a beamsize (FWHM)
of $7.4'$ arc at 18 cm. We need to find a phase reference within $3.7'$
of our target source. Reasoning like this reveals a misconception which,
though understandable in some respects, is nonetheless fundamentally
wrong. In the following paragraphs I will try to explain why.

Before launching into a more detailed discussion, I would first like
to state some facts about radio telescopes which most readers are
probably familiar with. There are two parameters which characterize
the sensitivity of a telescope: the effective collecting area, and
the system temperature. The effective area for a reflector is, of
course, the physical area multiplied by the efficiency (which takes
into account the surface accuracy, leakage if the surface is not
solid, and the feed illumination pattern). For a point source of
strength $S$, the effective area $A_e$ comes into the well-known
formula, $S=2kT/A$, whereby the ratio, $S/T$, which tells us the
source flux density required to raise the receiver temperature by
1 K (usually expressed in Jy/K), is seen to be determined by $A_e$.
We can then derive the system noise in Jy (SEFD) from the system
temperature. The receiver bandwidth can then be used to calculate
the noise level after a certain integration time. (All of the above
can be found in standard references on antennas and radio astronomy,
such as Kraus (\cite{kraus}).)

There are two other simple relationships for a two-element
interferometer which we might want to remember. Suppose that the
elements, labelled 1 and 2, have collecting areas $A_1$ and $A_2$,
and system temperatures $T_1$ and $T_2$. Then the interferometer
response (ignoring the individual element responses) for each of
these quantities will simply be the geometrical mean of element
1 and 2. For the interferometer collecting area we have,
$$ A_{int}=(A_1 A_2)^{1/2} $$
And for the combined system temperature,
$$ T_{int}=(T_1 T_2)^{1/2} $$
It should become clear later on why this is so.

\section{Overview of the EVN elements}

The EVN consists of ten regular stations, often augmented with a
handful of other ones. The regular station elements range in diameter
from 100 m to 25 m. An exceptional case is the Westerbork Telescope
array, which consists of fourteen 25 m dishes on a 2.7 km east-west
baseline. When phased-up and added together, it has the equivalent of
a roughly 90 m dish. Fewer Westerbork dishes can also be added (if
just twelve are included, the array length is decreased to 1400 m
and the effective diameter is about 85 m). Westerbork can also be
used as a 25 m single dish (Wb(1)). All of the stations which
participate in the EVN are listed in Table 1.

   \begin{table}
      \caption[]{Diameters of EVN elements.}
         \label{DiaEVN}
     $$ 
         \begin{array}{p{0.5\linewidth}l}
            \hline\hline
            \noalign{\smallskip}
            Diameter    &  \mathrm{Station(s)}  \\
            \noalign{\smallskip}
            \hline
            \noalign{\smallskip}
            200 m & \mathrm{Ar} \\
            100 m & \mathrm{Eb} \\
            90 m & \mathrm{Wb} \\
            76 m & \mathrm{Jb1} \\
            70 m & \mathrm{Rb70} \\
            32 m & \mathrm{Cm, Mc, Nt, Tr, Rb34} \\
            25 m/85 ft & \mathrm{Hh, Jb2, On-85, Sh, Ur, Wb(1)} \\
            20 m & \mathrm{On-60, Wz} \\
            14 m & \mathrm{Mh, Yb} \\
            \noalign{\smallskip}
            \hline
         \end{array}
     $$ 
   \end{table}

The Arecibo primary is 305 m (1000 ft) in diameter, but for normal
observations about 188 m of this is illuminated (Heiles et al.\ 
\cite{heiles}; the measured FWHM suggests a larger value). Wz mainly
participates in geodetic observations (and only operates at S/X-band).
The other small elements are mainly for high frequency observing.

\section{Response of a single dish}

Let us begin by considering the response of a single dish, as that
will provide us with some fundamentals before we turn to the case
of an interferometer. The antenna response, $A$, can be found by
taking the Fourier transformation (FT) of the autocorrelation of the
aperture illumination, $v$. For simplicity I will take a
one-dimensional case, with $u$ the coordinate in the aperture
plane ($u$ expressed in wavelengths), and $x$ the coordinate in
the sky or antenna beam frame (in radians). The autocorrelation
of $v(u)$ has the usual definition, $a(u)=\int v(l)\,v(u+l)\;dl$, or
in simplified notation, $a=v*v$. (I will assume that the telescope
response is symmetrical, and ignore the difference between
correlation and convolution.)

The Fourier transformations of $a$ and $v$ will, as is the usual
convention, be denoted by $A$ and $V$, respectively. The FT of $a$
has the usual definition: $A(x)=\int a(u)\,\exp(-2\pi i\,ux)\;du$. In
shorthand notation, $a(u) \rightarrow A(x)$. We want to
determine $A(x)$, and we have the following relationship:
$$ v*v=a\rightarrow A $$
We can use the convolution theorem (the FT of the convolution of
two functions is the product of their individual convolutions),
to get:
$$ A=V\times V $$
So, $A(x)=V^2(x)$, where $A(x)$ is the power polar diagram, while
$V(x)$ is the voltage polar diagram.

\section{Primary beam envelope of an interferometer}

What happens in the case of an interferometer? Then for our $v(u)$
we have to take $v_1(u)$ and $v_2(u)$ for our two elements 1 and 2,
and separate them by some distance, $\Delta u$ (= interferometer
baseline). This gives us a new aperture illumination, $v'(u)=v_1(u)+
v_2(u+\Delta u)$, so the autocorrelation we want is, $(v_1(u)+
v_2(u+\Delta u))*(v_1(u)+v_2(u+\Delta u))$. However, we are only
interested in the beam response envelope, so we will ignore the
autocorrelations ($v_1*v_1$, etc.), and assume that our elements
1 and 2 coincide ($\Delta u=0$), to get the cross-correlation of
the element illumination patterns, $a_{12}=v_1*v_2$. It is the FT
of this that we want:
$$ a_{12}\rightarrow A_{12} $$
and as in the case of the single dish, we get
$$ A_{12}=V_1\times V_2 $$
The combined beam envelope response is simply the voltage polar
diagram of one multiplied by that of the other.

The consequences of this can be most simply seen by considering a
limiting case. Suppose that the elements of our interferometer, with
diameters $D_1$ and $D_2$, are very unequal with $D_1>>D_2$. Then
their beam sizes ($\theta$) will also be unequal, $\theta_1<<\theta_2$.
Now, $\theta_2$ large means $V_2(x)\simeq$ const $=V_0$, which gives,
$$ A_{12}(x)\simeq V_1(x)\times V_0 $$
and the width of the combined interferometer beam will be determined
by the {\it voltage} pattern, $V_1$. For a (nearly) gaussian-shaped
beam, the FWHM response will be about 40\% larger (greater by $\sqrt{2}$
for a true gaussian), and the beam area will of course be doubled. In
reality, the illumination pattern (even if gaussian) will be truncated
at the edge of the aperture, leading to sidelobes and nulls in the
antenna pattern. The nulls will, naturally, have the same location in
the voltage beam, so that will not be changed in the interferometer
response. The relative strength of the sidelobes will be increased.

\section{Primary beam envelope of EVN interferometers}

As we saw in Section 2, the EVN station elements differ considerably
in size (Table 1). Here I would like to consider the consequences of
these differences on the combined beam envelope for some common
baseline configurations. The calculations have been done for a
wavelength of 18 cm, as this is widely used in the EVN and encompasses
all stations. For Arecibo, the beamwidth (FWHM) measured by Heiles et
al. (2001) at 1666 MHz has been used. For Effelsberg, the value
measured by Reich et al.\ (1978) at 2700 MHz has been scaled to 1666
MHz. For the other elements, the Effelsberg value has been scaled by
the dish diameter. Table 2, most of which is self-explanatory,
gives the results for most baseline combinations of elements ranging
from Arecibo and Effelsberg, to the 32 and 25 m dishes. The fourth
column, headed $\theta_{12}$, is the combined interferometer beam
envelope for the elements in column 1. The fifth column, $\Omega_1/
\Omega_{12}$, is the factor by which the FWHM interferometer beam
area is greater than the single-dish beam area of the larger element.

   \begin{table}
      \caption[]{EVN element and interferometer beam properties at 18 cm.}
         \label{BemEVN}
\centering                          
\begin{tabular}{c l l l l}        
\hline\hline                 
            \noalign{\smallskip}
   $\rm Sta_1 * Sta_2$ & $\theta_1$ & $\theta_2$ & $\theta_{12}$ &
 $\Omega_1/\Omega_{12}$ \\
            \noalign{\smallskip}
            \hline
            \noalign{\smallskip}
            Ar * Eb    & $2.9'$ arc & $7.4'$ arc & $3.8'$ arc & 1.74 \\
            Ar * Jb1   & $2.9'$ &    $10'$       & $3.9'$     & 1.84 \\
            Ar * 32 m  & $2.9'$ &    $23.2'$     & $4.1'$     & 1.97 \\
            Ar * 25 m  & $2.9'$ &    $30'$       & $4.1'$     & 1.98 \\
            Eb * Jb1   & $7.4'$ &    $10'$       & $8.4'$     & 1.3 \\
            Eb * 32 m  & $7.4'$ &    $23.2'$     & $10'$     & 1.83 \\
            Eb * 25 m  & $7.4'$ &    $30'$       & $10.2'$     & 1.9 \\
            Jb1 * 25 m & $10'$  &    $30'$       & $13.4'$     & 1.8 \\
            32 m * 25 m & $23.2'$ &  $30'$       & $26.0'$     & 1.25 \\
            \noalign{\smallskip}
            \hline
\end{tabular}
\end{table}

The increase in beam area for the interferometer pair over that of the
larger element is naturally greater when the elements have very different
diameters. The Arecibo and Effelsberg combinations with the smaller
elements illustrate this quite clearly. There is, of course, no sharp
``edge'' to the primary beam, so one has considerable discretion in
deciding where to draw the line, but even taking it at $2/\pi$ or $1/e$
or whatever one's favorite cutoff might be, within reason, similar ratios
will hold. The Westerbork array has not been considered here, as it is
a rather exceptional case. But for it too, the width of the fan beam
at 18 cm will increase from about $20''$ arc to $28''$ arc. However a
full treatment is beyond the scope of this short note.

Users of arrays with equal elements, like the VLBA, do not have to
concern themselves with these matters, except in the case of global
experiments (but then, that's no longer the VLBA as such, nor are
the elements equal).

\section{So what?}

Most VLBI users are only concerned with emission within the inner
$<1''$ arc of their pointing direction. For them, the points raised
above may be of no
more than academic interest. However, for those wanting to pursue the
structure of numerous sources in a larger region of the sky, it is
worth considering the correct (combined) beam size. The area
covered (and hence the number of sources) can be nearly double
that of the larger element used as a single dish. This having been
said, most of the beam sizes in Table 2 extend well beyond the distance
where bandwidth decorrelation and integration time smearing will
normally limit
effective imaging. The most likely consequence of the beamsizes lies
in the realm of using in-beam phase reference sources. Doubling the
beam area means doubling the chance of finding a suitable reference.

There may, in the future, be better prospects for imaging over a
larger portion of the primary beam envelope. With correlators and
storage devices able to handle larger data volumes, the possibilities
for producing many frequency channels with short integration times
will also increase. From the observational side, there is increasing
pressure for wider fields of view. This is aptly illustrated by
recent EVN observations of the Hubble Deep Field (Garrett et al.\
2001). Similarly, the combination of EVN and MERLIN also tends to
move one into the realm of wide(r)-field mapping.

\section{Conclusions}

The results presented here on the primary beam of an interferometer
with unequal elements should be familiar to anyone working in the
field of radio interferometry. Unfortunately, my experience has been
that these facts are often overlooked. In stating the obvious, I hope
that I have not bored too many readers.

\begin{acknowledgements}
I thank various colleagues for discussions held on this and realted topics
over the past few years. ASTRON is funded by the Netherlands Organization
for Scientific Research (NWO).
\end{acknowledgements}

\cleardoublepage

\end{document}